\documentclass[twocolumn]{aastex631}

\usepackage{amsmath}
\usepackage{threeparttable}
\usepackage{graphicx}        
\usepackage{CJK}
\usepackage{tablefootnote} 
\usepackage{threeparttable}
\usepackage{longtable}
\usepackage{lipsum} 
\usepackage{threeparttablex}

\usepackage{times}
\usepackage{amsmath}
\usepackage{amssymb}
\usepackage{xspace}
\usepackage{mathrsfs} 
\usepackage{tabularx}
\usepackage{fixltx2e}
\usepackage{color}
\usepackage{placeins}
\usepackage{comment}

\shorttitle{Where is the Super-Virial Gas?}
\shortauthors{Roy et al.}

\begin{document}
\UseRawInputEncoding
\title{Where is the Super-virial Gas? II: Insight from the Survey of Galactic Sightlines}

\correspondingauthor{Manami Roy}
\email{roy.516@osu.edu}
\author[0000-0001-9567-8807]{Manami Roy}
\affiliation{Center for Cosmology and Astro Particle Physics (CCAPP), The Ohio State University, 191 W. Woodruff Avenue, Columbus, OH 43210, USA}
\affiliation{Department of Astronomy, The Ohio State University, 140 W. 18th Ave., Columbus, OH 43210, USA}
\author[0000-0002-4822-3559]{Smita Mathur}
\affiliation{Center for Cosmology and Astro Particle Physics (CCAPP), The Ohio State University, 191 W. Woodruff Avenue, Columbus, OH 43210, USA}
\affiliation{Department of Astronomy, The Ohio State University, 140 W. 18th Ave., Columbus, OH 43210, USA}
\author[0000-0002-9069-7061]{Sanskriti Das}
\altaffiliation{Hubble Fellow}
\affil{Kavli Institute for Particle Astrophysics and Cosmology, Stanford University, 452 Lomita Mall, Stanford, CA 94305, USA}
\author[0000-0001-6995-2366]{Armando Lara-DI}
\affil{Instituto de Astronomia, Universidad Nacional Autonoma de Mexico, 04510 Mexico City, Mexico}
\author[0000-0001-6291-5239]{Yair Krongold}
\affil{Instituto de Astronomia, Universidad Nacional Autonoma de Mexico, 04510 Mexico City, Mexico}
\author{Anjali Gupta}
\affil{Columbus State Community College, 550 E Spring Street, Columbus, OH 43210, USA}

\begin{abstract} \label{abstract}
Recent observations have revealed a super-virial temperature gas phase at 
log(T/K) $\sim7$ in the Milky Way, challenging existing galaxy-formation models. This hot gas phase was discovered toward extragalactic absorption sightlines and blank-sky emission fields, both at high galactic latitudes. The location of this hot component is unknown; is it in the extended circumgalactic medium (CGM) or in the interstellar medium (ISM) instead? We analyzed X-ray spectra from Chandra's High-Energy Transmission Grating (HETG) observations of 27 Galactic X-ray binaries (XRBs) to investigate whether the hot gas component is present in the ISM. We searched for absorption lines of SXVI K$\alpha$, SiXIV K$\alpha$, and NeX K$\alpha$, which are the tell-tale signatures of the hot gas and which have been detected toward extragalactic sightlines. Of the 27 targets, these lines were detected in the spectra of only 7, with two sources displaying broad line features likely intrinsic to the XRB systems. 
Additionally, most of the detected lines are time-variable, reinforcing their likely association with the XRBs. Our results suggest that the super-virial temperature gas is not a widespread component of the ISM but may instead be located in extraplanar regions or the extended CGM, {which aligns} with some recent simulation results. 
 \end{abstract}


\section{Introduction}
Galaxies, the fundamental building blocks of our universe, have primary components such as a star-forming disk and the Interstellar Medium (ISM), a complex, multi-phase environment within the disc containing a mix of dust, molecules, and atoms. Above and below the galactic disk lies a more diffuse and ionized region compared to the ISM, known as Circumgalactic Medium (CGM) [For a review see \cite{Tumlinson2017}, \cite{Mathur2022}]. 

Detections of SiII, SiIII, SiIV, NeIX, OVI, and OVII
along with neutral atomic and molecular hydrogen from the Milky Way's ISM have revealed its diverse composition, encompassing various gas phases. These elements can be ionized and dispersed by energetic events such as stellar winds, supernova explosions, and interactions within binary star systems throughout their surroundings. Also, observations from the COS spectrograph on the Hubble Space Telescope, as well as from XMM-Newton and Chandra, have revealed the multiphase nature of the CGM gas around Milky Way and Milky Way-like galaxies in both UV [e.g. \cite{sembach2003highly, Tumlinson2011, Werk2016, Richter2017}] and X-ray [e.g. \cite{Nicastro2002, Williams2005, Henley2010, Gupta2012, Fang2015, 2021ApJ...918...83D}]. In contrast to the Galactic disk, where energetic processes account for the observed ionized elements in the ISM, the CGM does not have a well-defined mechanism for producing and ionizing gas. Numerical simulations suggest that feedback processes from star-formation activities and active galactic nuclei, which expel ionized metals from the disk and the nucleus into the CGM, play a significant role in shaping the CGM.
\begin{figure}
\includegraphics[width=1.0\linewidth]{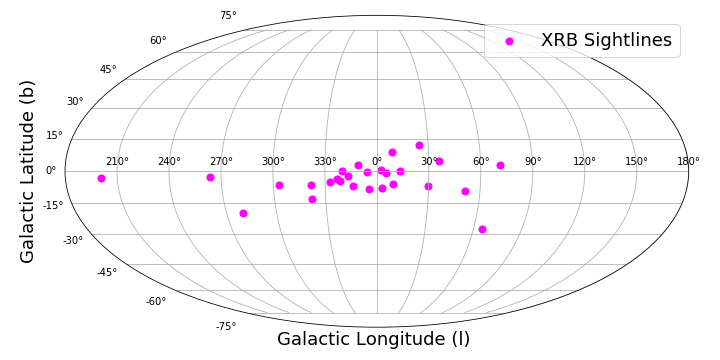}
\caption{Skyplot of the sample of $27$ XRBs we analyze in this work, showing their Galactic latitude and longitude.}
\label{f:co}
\end{figure} 

\begin{figure*}
\includegraphics[width=0.33\textwidth]{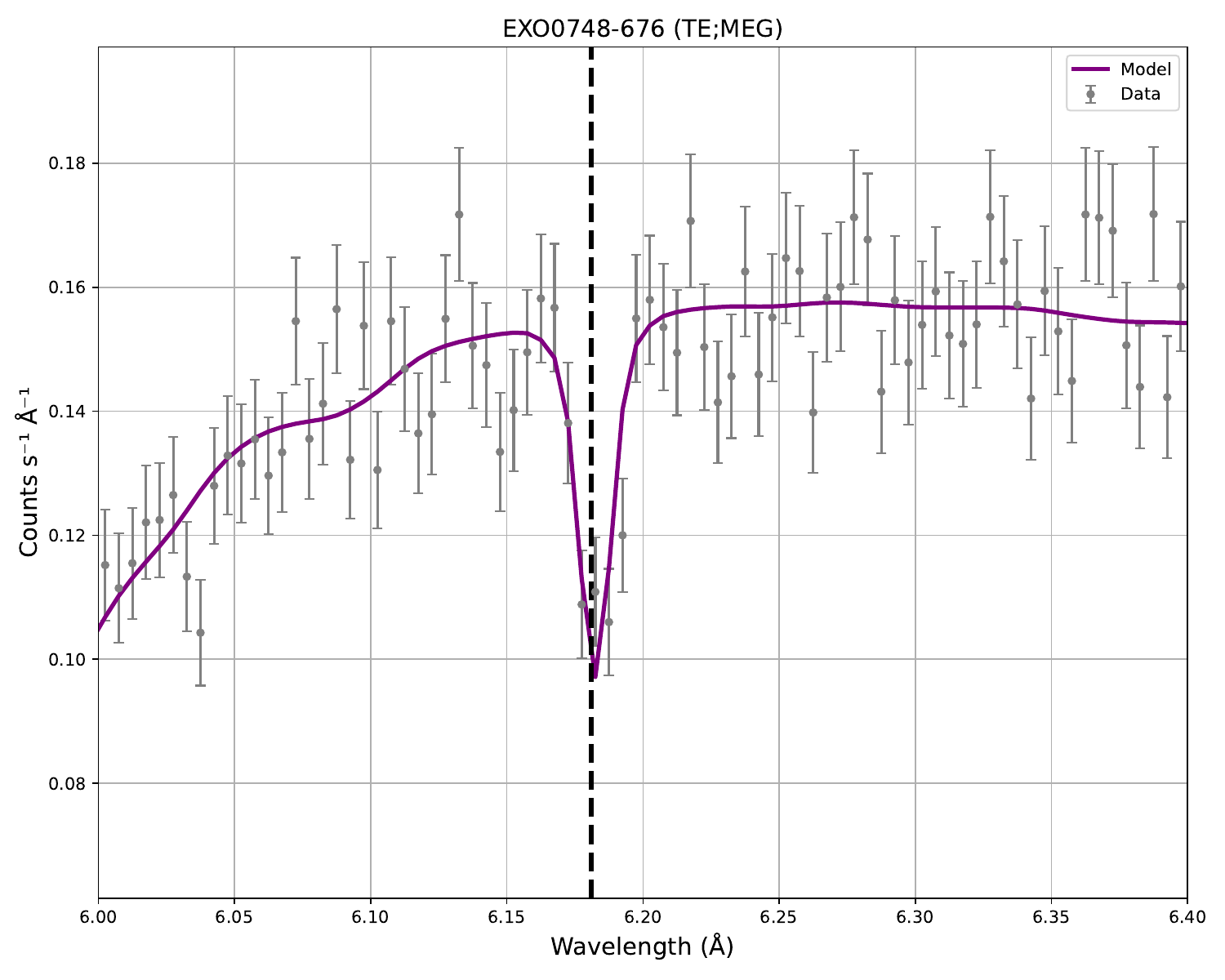}
\includegraphics[width=0.33\textwidth]{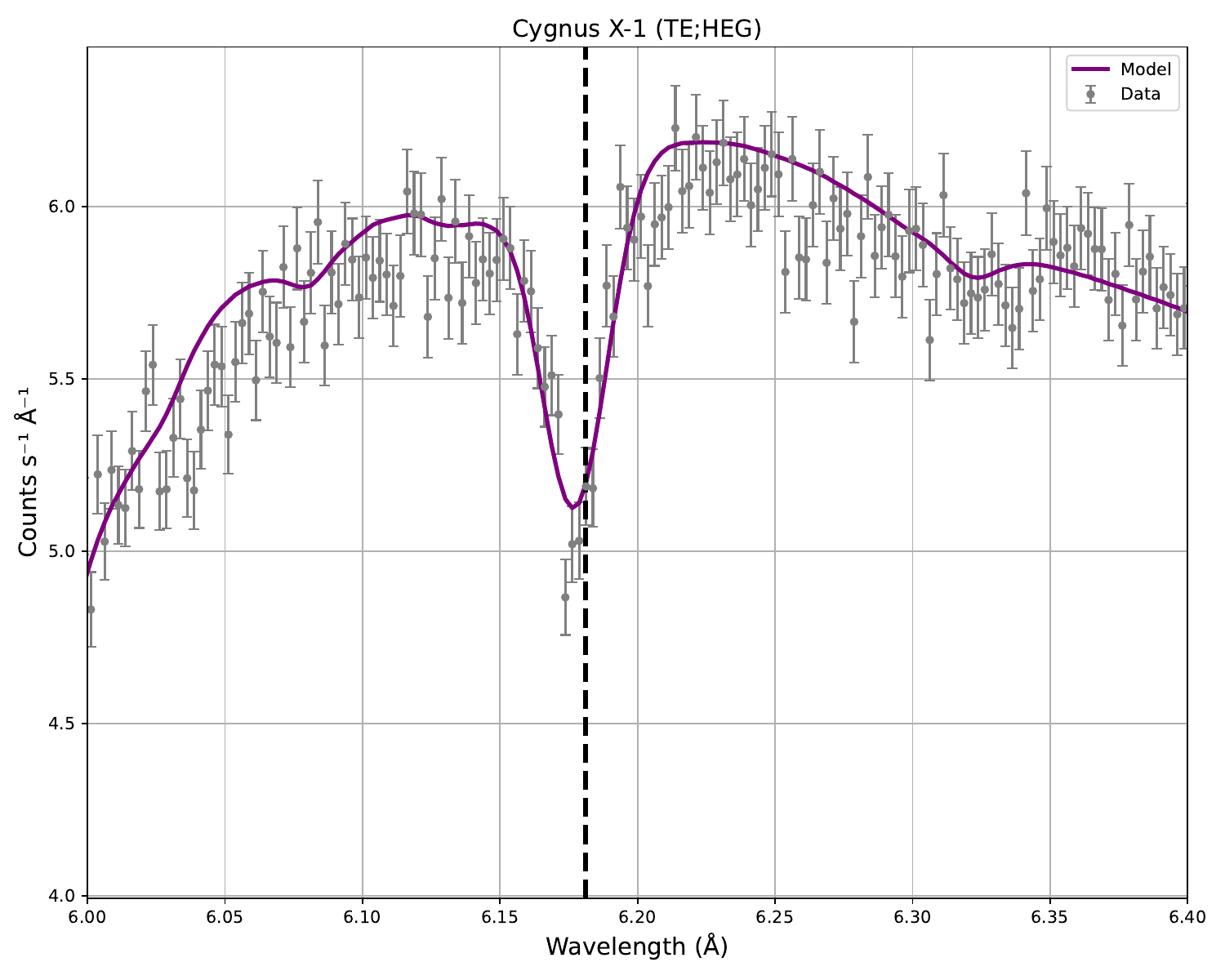}
\includegraphics[width=0.33\textwidth]{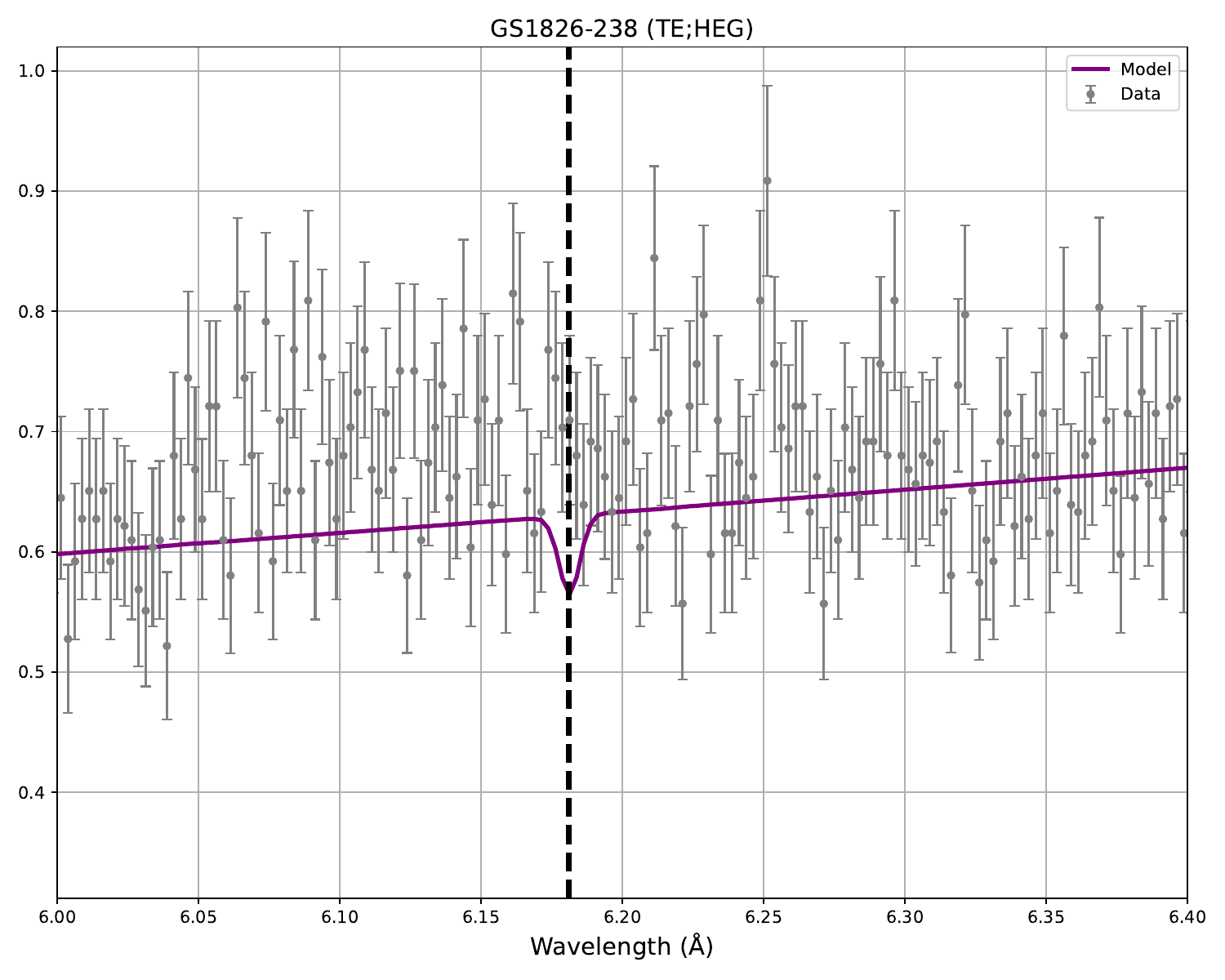}
\caption{Examples of SiXIV K$\alpha$ line profiles observed in three XRB sources. The black points with error bars represent the data, which are fit using Power-law and Gaussian models (shown in solid purple lines): a narrow Gaussian (left panel), a broad Gaussian (middle panel), and a 3$\sigma$ upper limit(right panel). 
\\ \textbf{Summary:} The narrow and broad Gaussian fits capture 3$\sigma$ detected SiXIV line, while the 3$\sigma$ upper limit represents cases with no significant line detection. The vertical dashed lines indicate the expected rest wavelength of the SiXIV transition.}
\label{lines}
\end{figure*} 
Based on these observations, the CGM was thought to have three phases; virial warm-hot gas (log(T)$\sim 6-6.5$), warm gas (log(T)$\sim 5-5.5$), and cool gas (log(T)$<4$). However, this understanding was shaken with the recent discovery of the super-virial component (log(T)$\sim 7-7.5$ \citep{2019ApJ...882L..23D}) in the Milky Way CGM using absorption lines from XMM-Newton observations towards the blazar IES 1553+113. Since then, multiple detections of this hotter super-virial gas have been made in both emission and absorption [e.g. \cite{2019ApJ...887..257D, Bluem, 2023NatAs...7..799G, 2023ApJ...952...41B, 2023ApJ...946...55L, 2024MNRAS.527.5093M}]. For example, \cite{2019ApJ...887..257D} detected the super-virial component in emission along the same sight-line of IES 1553+113 using XMM-Newton. \cite{Gupta2021} identified this component and the virial component across four sightlines in emission using Suzaku data. Later, \cite{2023ApJ...952...41B} also observed this super-virial gas in the Milky Way's CGM in emission towards the blazar Mrk 421 and five other nearby sightlines. In a recent study, \cite{2024MNRAS.527.5093M} detected NeX line, associated with super-virial temperature gas, along the sightline to NGC\,3783 and \cite{2023ApJ...946...55L} used stacked X-ray spectral observations toward multiple extragalactic sightlines to detect SiXIV K$\alpha$ and SXVI K$\alpha$ lines, signature of the super-virial temperature gas in the Milky Way's CGM. Additionally, it was found that the super-virial temperature hot component is not limited to specific sightlines and is detected all across the sky by all-sky surveys in both absorption \citep{2023ApJ...946...55L,Lara2024} and emission \citep{2023NatAs...7..799G, Bluem, Ponti}.

For the Milky Way mass galaxy, the virial component of the CGM gas is predicted by galaxy formation simulations, but the gas at log(T/K)$\sim7-7.5$ was not predicted. It raises several important questions about the origin and location of the super-virial gas. Is it truly from the CGM, or is it in the ISM or the extra-planar region outside the ISM?  To answer these questions, we examine Chandra's High-Energy Transmission Grating (HETG) X-ray observations of Galactic X-ray binaries (XRBs). In the pilot study, \cite{Lara2024} analyzed three XRB sightlines and did not detect the super-virial hot gas. However, to conclude that this phase is not present in the ISM it is important to study a large sample with all-sky coverage and better statistics.  Therefore,  we survey a large extended sample of 27 XRBs and look for the NeX, SiXIV, and SXVI absorption lines, which are the tell-tale signs of the super-virial hot component and which have been previously detected along extragalactic sightlines. This experiment is similar to that done by \cite{Nic2016} to separate the ISM and CGM components of the virial-temperature gas traced by the OVII absorption line ($\lambda=21.602$\AA). These authors found that the OVII-bearing gas permeates both the ISM and the CGM, but the column density in extragalactic sightlines is dominated by the CGM. 

This paper is organized as follows: Section \ref{S: data} details the data sample, while Section \ref{S: ana} outlines the data analysis methodology. In Section \ref{S: result}, we present our findings, followed by a discussion of their implications in Section \ref{S: discussion}. Finally, our conclusions are summarized in Section \ref{S: conclusion}.
\begin{figure*}
\includegraphics[width=0.49\textwidth]{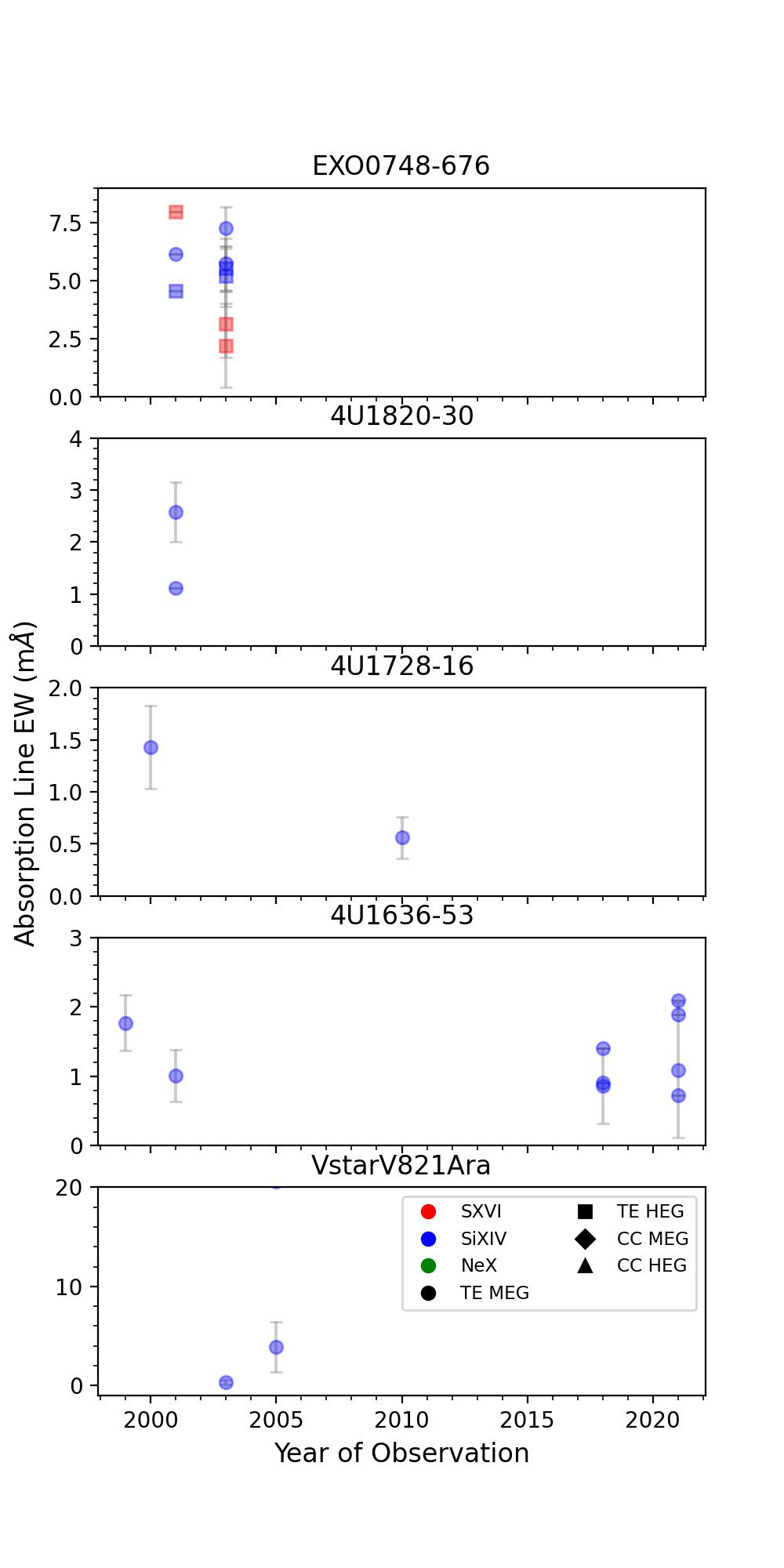}
\includegraphics[width=0.49\textwidth]{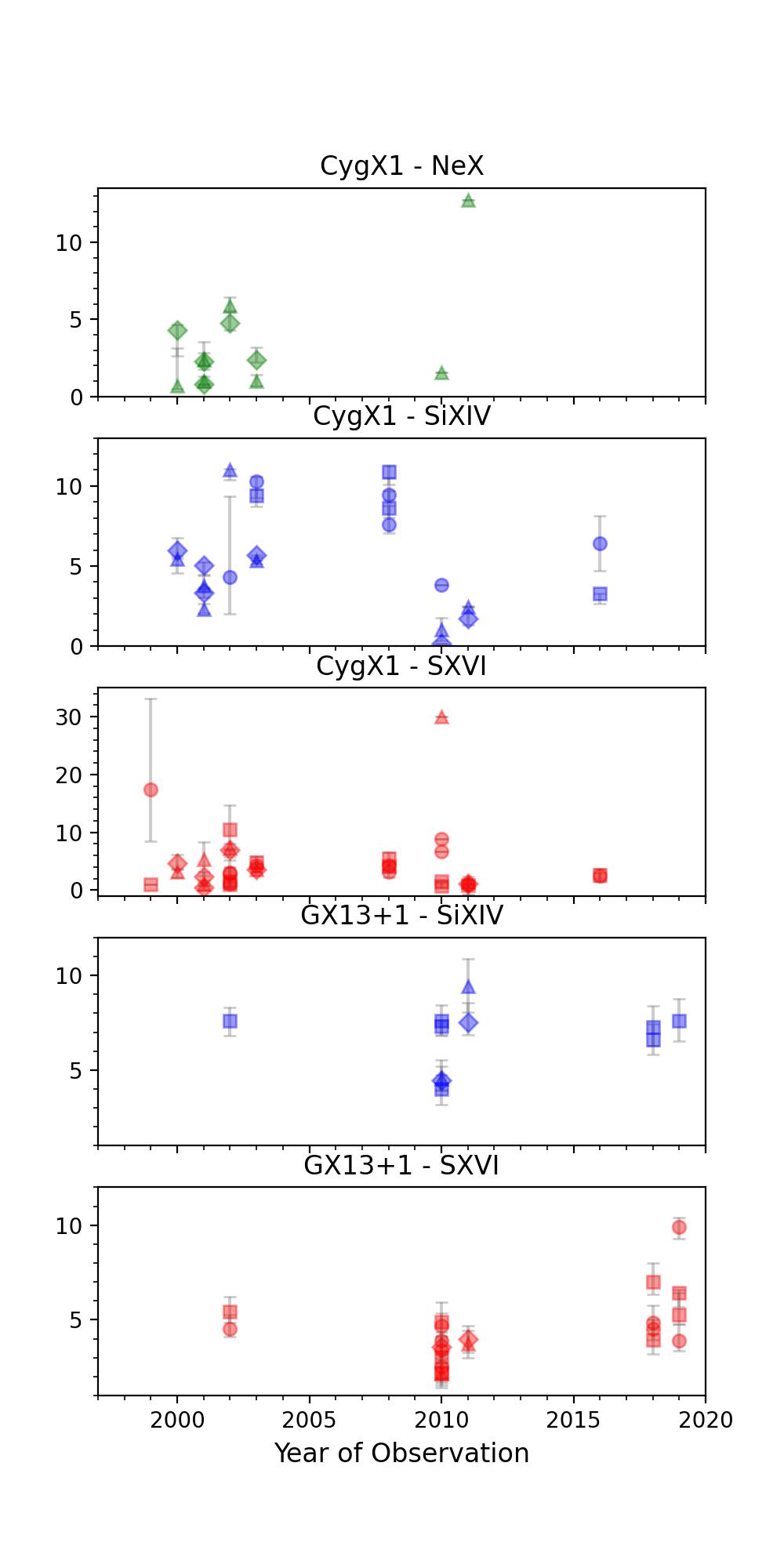}
\caption{Temporal variation of NeX, SiXIV, and SXVI lines detected at the 3$\sigma$ significance level across different XRB sources, showing the 1$\sigma$ errors and 1$\sigma$ upper limits. The observed lines exhibit clear variability over time, except for SXVI in EXO 0748-676 sightline, where the variability remains inconclusive due to a limited number of data points and a short observational time gap. The data points without error bars indicate the 1-$\sigma$ upper limit.\\
\textbf{Summary:} The variability of these detected lines indicates that these lines are most likely originating from variable XRB sources, not from diffuse ISM.}
\label{f:var}
\end{figure*} 

\section{data} \label{S: data}
We obtained archival Chandra ACIS-S HETG observations for 27 XRBs, which were taken from the sample used by \cite{Nicastro2016, Gatuzz2021}. We show the location of our XRB sample in terms of Galactic latitude and longitude in Figure \ref{f:co}. This dataset includes approximately 190 observations, with a total combined exposure time of 5758 ks (see Table \ref{tab:source_description}). 

The observations were reprocessed using the `\textit{chandra\_repro}' script within the Chandra Interactive Analysis of Observations (CIAO) software, version 4.13. We then combined the spectra using the `combine\_grating\_spectra' command in CIAO, resulting in 4 stacked spectra for each XRB: one for the High Energy Gratings (HEG) continuous clocking (CC) mode observations, another for the HEG timed exposure (TE) mode observations, a third for the Medium Energy Gratings (MEG) CC observations, and a final one for the MEG TE observations. Thus we created $27\times 4$ spectra which were analysed as discussed in \S3. The MEG has a resolving power ($\lambda$/$\Delta \lambda$) of 1000 at 12.4 \text{\AA}, while the HEG has an $\lambda$/$\Delta \lambda$ of 660 at 15 \text{\AA}.


In the 4-19 \text{\AA} spectral range, TE observations can be significantly affected by pileup when the sources are in a high flux state (e.g., \citealt{Roga2021}). Pileup occurs when multiple photons are recorded as a single event, resulting in information loss and spectral distortion. {However, pileup minimally impacts narrow absorption features from grating spectra, as described by \cite{Roga2021}.  Grating spectra mitigate the effects of pileup due to the fundamental mechanism of photon dispersion. Moreover, as the narrow absorption lines are confined to specific energy ranges, they correspond to distinct spatial locations on the detector in grating mode.  Therefore, the likelihood of multiple photons arriving simultaneously at those specific energies corresponding to these narrow features is low.  
Additionally,  the continuum distortion caused by pileup, where low-energy photons are shifted to higher energies, is minimized in grating mode. The dispersion spreads the continuum flux over a larger area, making the effects of pileup less concentrated. As a result, the contrast between the absorption line and the continuum remains largely intact, enhancing the detectability of weak or narrow lines. }Therefore, we analyzed both TE and CC spectra for all these targets, focusing on the narrow absorption features.
\section{analysis} \label{S: ana}

For our spectral fitting, we employed the XSPEC software (v12.13.0) and used chi-squared ($\chi^2$) statistics to analyze the CC MEG and HEG spectra of our sample XRBs with a high signal-to-noise ratio of $>50$. We conducted a similar analysis for the TE observations.

Following \cite{Lara2024}, our spectral modeling focused on the region within $\pm 0.25$ \AA\ around the rest wavelengths of the ionic transitions of interest: SXVI K$\alpha$ ($\lambda = 4.729$ \AA), SiXIV K$\alpha$ ($\lambda = 6.182$ \AA), and NeX K$\alpha$ ($\lambda = 12.134$ \AA).

Initially, we fitted the local continuum within these narrow spectral ranges using a power-law model (XSPEC model \textit{pow}).  The power-law models for the different spectral regions were allowed to vary independently. If additional spectral features were present, we included a Gaussian profile (XSPEC model \textit{agauss}) to account for them.

Next, we modeled the potential presence of absorption features with a narrow Gaussian profile centered at the rest wavelengths of SXVI K$\alpha$, SiXIV K$\alpha$, and NeX K$\alpha$. The position of each Gaussian profile was set to the expected line value, with an allowed variation of $\pm 0.012$ \AA\ in some cases to account for the resolution element of HEG and $\pm 0.023$ \AA\ for the resolution element of MEG. We fixed the line width to zero \AA\ (as the lines are unresolved at HETG resolution) and allowed the normalization to vary. In some cases, we also fit with a broad Gaussian line (with non-zero line width) or two narrow Gaussian lines, as required by the data (see \S 4).

\section{results} \label{S: result}
For the 27 XRBs, we have $27\times4 =108$ spectra (HEG+MEG+CC+TE) and $108\times 3=324$ possible detections for the three lines (SXVI K$\alpha$, SiXIV K$\alpha$, and NeX K$\alpha$). We detected only 25 lines with $3\sigma$ or better significance:  2 of Ne X, 14 of Si XIV, and 9 of S XVI. All the detected lines were in the spectra of 7 XRBs (out of 27), as shown in Figure \ref{lines} and table \ref{fit}. 

Of the 25 detected lines, 16 appeared to be resolved. Therefore, we fitted them with broad Gaussian profiles or with two narrow Gaussian (marked by double stars in the table where we quoted the values for the fit with a broad Gaussian). All the results are summarized in Table \ref{fit}, where the columns (left to right) are the names of the sources, exposure mode, grating name, and line equivalent width. The errors quoted are $1\sigma$ and the bold numbers are the lines where we got at least $3\sigma$ detections. The values without error bars are $1\sigma$ upper limits. In Figure \ref{lines}, we show examples of SiXIV line profiles toward three XRB sightlines (with a narrow Gaussian in the left panel, a broad Gaussian in the middle panel, and a $3\sigma$ upper limit in the right panel). The narrow and broad Gaussian fits show the $3\sigma$ detections of the SiXIV line, while the $3\sigma$ upper limit represents cases with no significant line detection. The vertical dashed lines indicate the expected rest wavelength of the SiXIV transition.
\subsection{Detection or not?} Among all the detections, for four of the sources, we noticed that there are detections of SiXIV in MEG but not in HEG. HEG has a lower effective area in the relevant energy range than MEG, but the upper limits of HEG are lower than MEG detections. Therefore, we investigate these cases further. We simultaneously fit the spectra from HEG and MEG and look for the SiXIV line. We summarize our finding in Table \ref{tab:simul}, where the third column denotes EW of individual fitting, and the fourth column denotes EW of simultaneous fitting of HEG and MEG spectra. As anticipated, the EWs of all of these four detections have decreased and now the line is detected in both HEG and MEG. However, the detection significance has decreased from more than 3$\sigma$ to close to 3$\sigma$.

\begin{table}
    \centering
    \begin{tabular}{|c|c|c|}
        \hline
        \textbf{Source} & \textbf{Grating} &  \textbf{EW$_{\rm sim}$} (m\AA) \\
        \hline
        4U1636 [TE] & HEG &   -0.37$^{+0.13}_{-0.14}$  \\
        & MEG & -0.42$^{+0.16}_{-0.15}$  \\
        \hline
        4U1728-16 (GX9+9) [TE]  & HEG & -0.34$^{+0.11}_{-0.11}$  \\
        & MEG &  -0.41$^{+0.14}_{-0.14}$ \\
        \hline
        V*V821Ara (GX 339-4) [TE] & HEG & -0.42$^{+0.15}_{-0.14}$ \\
        & MEG &  -0.48$^{+0.16}_{-0.17}$ \\
        \hline
        4U1820-30 [TE] & HEG & -0.76$^{+0.26}_{-0.26}$  \\
        & MEG & -0.99$^{+0.34}_{-0.34}$   \\
        \hline
    \end{tabular}
    \caption{Values of EW of SiXIV detections after simultaneous fitting of spectra from HEG and MEG}
    \label{tab:simul}
\end{table}

\section{discussion} \label{S: discussion}

Different studies have identified a warm-hot gas phase in the Milky Way's ISM with a temperature around log(T /K) $\sim6$ (e.g. \citep{Nicastro2016}). NeIX originates from this gas phase and has been detected in the sightlines analyzed in these studies. In contrast, ions like NeX, SiXIV, and SXVI, which indicate even hotter gas, have not been observed in the spectra of most of these XRBs. Additionally, there are no known detections of these higher charge states in the Galactic ISM in existing literature. This implies that while gas at log(T /K) $\sim6$ is commonly present in the ISM, the hotter gas is rare.
Therefore, the contribution of the ISM to the supervirial hot component observed in extragalactic sightlines is minimal at best and much lower than the contribution from the warm-hot component (e.g. 
\citealt{Nicastro2016}).

{Additionally, we compare the line detection limits for the Galactic XRB sightlines to those for the extragalactic lines of sight. For EXO0748-676, the EW limits for SiXIV are higher and for SXVI they are comparable with the extra-galactic sightlines [comparing EW from table 2 \cite{2023ApJ...946...55L}]. For other XRB sightlines, narrow line EW limits are $\sim30-50$ times lower than extragalactic sightlines.  Typically, the detection limits for Galactic X-ray binary sightlines are lower (i.e., weaker lines can be detected) becuase XRBs are bright X-ray sources. Extragalactic sources, on the other hand, are fainter, requiring much longer observation times to detect comparably weak lines.}

Next, let us discuss the line detections and analyze whether they arise in the  ISM or are intrinsic to the XRB source. We have tried to answer this question in three ways, as discussed below. 
\begin{figure*} 
\includegraphics[width=\textwidth]{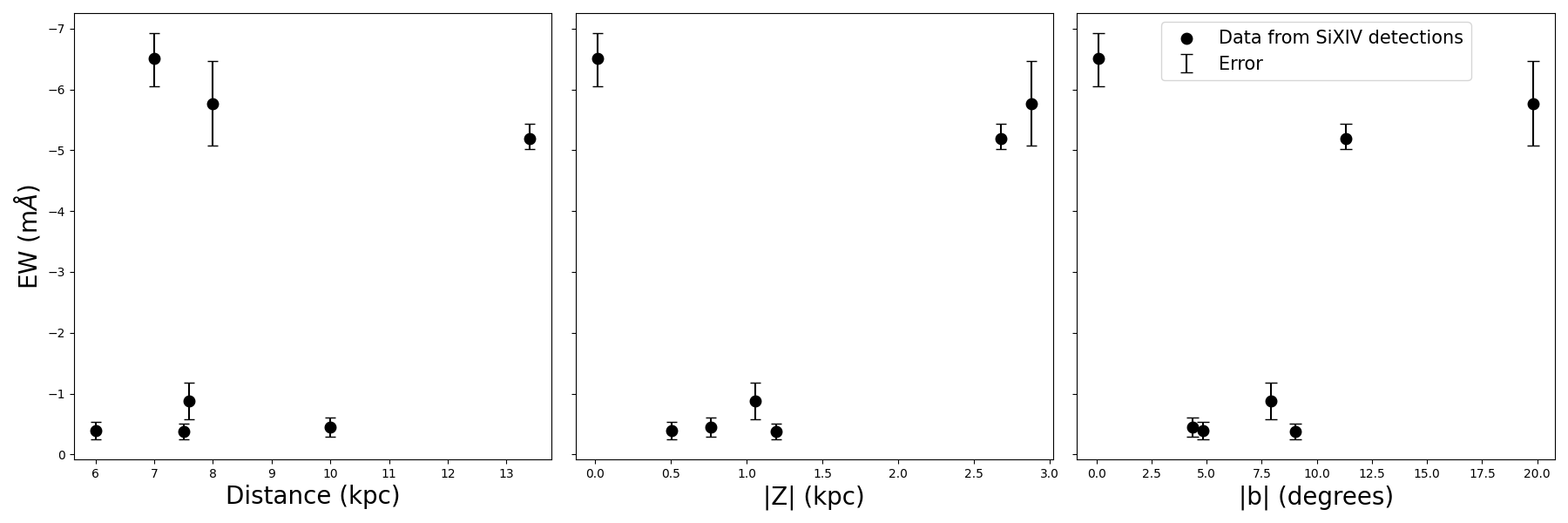}
\includegraphics[width=\textwidth]{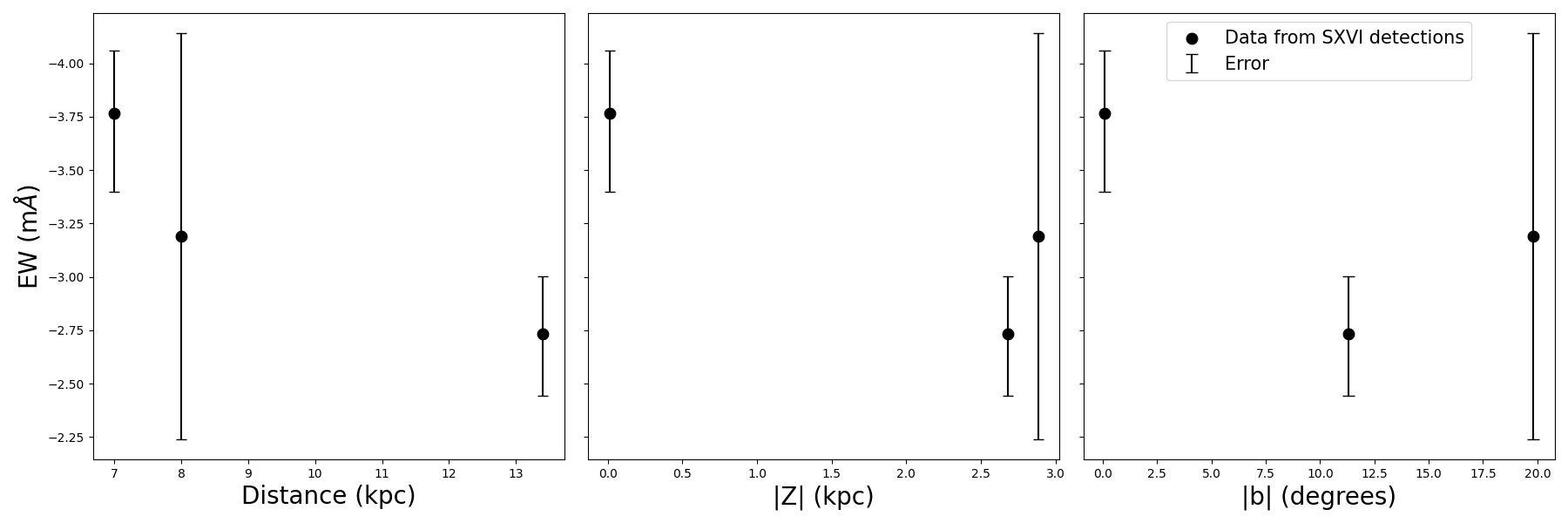}
\caption{The equivalent width (EW) of Si XIV (\textit{top}) in 7 XRB sightlines and S XVI (\textit{bottom}) lines in 3 XRB sightline as a function of distance from us (\textit{left}),  height above the Galactic plane (\textit{middle}), and Galactic latitude (\textit{right}). NeX has been detected only in one XRB sightline, therefore correlation study is not possible and is not shown here. 
\\
\textbf{Summary:} We find no clear correlation between EW and distance, height, and latitude. The lack of clear correlation suggests that these lines do not originate from the diffuse interstellar medium (ISM).}
\label{EW_r_Z}
\end{figure*} 

\subsection{ISM or intrinsic to XRB? : From Variability Test}
 We performed a variability test whenever we had a detection. We take individual observations made at different times and fit the same line which is detected in the stacked spectrum. We find that some of the detected lines are variable with time whereas some of them are not. We note the equivalent width of individual observations in the table \ref{t: var}. We also plotted the equivalent width with time of observations for the detected lines in Figure \ref{f:var} and summarized our finding in Table \ref{t: var}. In the left panel of Figure \ref{f:var}, we show 5 XRBs for which we can fit narrow lines and in the right panel we show the individual ions of two XRBs fitted with broad lines. We discuss all the cases here. 
 \begin{itemize}
 \item In the case SXVI line in EXO0748-676, the upper limit in 2001 is consistent with the 2003 detection and error bar. However, the two data points from 2003 do not coincide completely, which can be suggestive of variability. For the SiXVI line, the upper limit in 2001 is not consistent with the 2003 detections, 
 suggesting variability.  
 \item For 4U1820-30, the upper limit of SiXIV in 2001 is not consistent with the detection in the same year, indicating variability. 
 \item For the SiXIV in the 4U1728-16 spectra, we see a clear variability from the year 2000 to 2010. \item For VstarV821Ara, the upper limit of SIXIV in 2003 is not consistent with detection in 2005, suggesting variability.
 \item for CygX1, the upper limits, and detections vary with time for all the ions. 
  \item for GX13+1, the upper limits, and detections vary with time for all SiXIV and SXVI ions. 
\end{itemize}
If the lines are variable then they likely arise intrinsically in the XRB and if they are not variable then they might be intrinsic or might arise from the intervening medium. Therefore, as shown in Figure \ref{f:var}, most of the detected lines are variable and are most likely intrinsic to XRBs.

\subsection{ISM or intrinsic to XRB? : From line width}
The line width can also indicate whether lines are from the ISM or if they are intrinsic to the XRB source. For example, wind, and large turbulence in the XRB may give rise to a broad profile. Alternatively, however, turbulence in the ISM can also broaden the line. Two narrow lines at slightly different heliocentric velocities may overlap and may also mimic a broad line. 

For most of the detections, except for GX13+1 and CygX1, the lines were well-fitted with one narrow Gaussian line. However, for GX13+1 and CygX1, the lines are broad and we fitted them with a broad Gaussian or two narrow Gaussians. However, we can not choose between these two models based on the chi-squared value alone as reduced chi-square is similar for both the fit. However, \cite{Das2024} demonstrate that fitting two unresolved Gaussians with a single unresolved Gaussian does not significantly overestimate or underestimate the input EW, regardless of the LOS velocities and amplitudes, ruling out the possibility that our data represent purely thermally broadened multiple components appearing as a single non-thermally broadened component. 

Further, we measure the non-thermal broadening of these lines and find that in the case of CygX1 and GX13+1, the broad lines (SiXIV and SXVI) have Full-width half maximum (FWHM) of $\sim1000$ km/sec and $\sim1200$ km/sec respectively. 
The thermal FWHM of SiXIV and SXVI lines are 180 km/sec and 164 km/sec and the instrumental width is about $\sim500-700$ km/sec. Thus the thermal width is not high enough to be resolved by HETG, however, the non-thermal width is well-resolved and well above the instrumental width. Also, the non-thermal width is 2-5 times more than the thermal width, where broadening can rise from turbulence, wind from XRB, or turbulence in ISM.

{However, note that the presence of the binary-related absorption lines can affect the detection of the weak ISM lines. In order to assess this, we calculate the upper limit of the detection of a narrow line by fitting the observed line with a broad line and one narrow line. We find that the EW limit of the narrow line is $\sim30-50$ times smaller than the EW of the broad line. This implies that the narrow line can be detected in the presence of a broad line even if the EW of the narrow line is 30-50 times smaller than the broad line EW.}

\subsection{ISM or intrinsic to XRB? : From correlation with Galactic distance and height}
Another way to check whether the lines are arising in the intervening medium is to investigate the correlation between the EW of the detected lines and the height from the Galactic plane, distance from us, and Galactic latitude. If the lines arise in the diffuse ISM then the EW will increase with increasing distance and decreasing height and Galactic latitude, as we probe more of the disk, hence more of the ISM. We plotted the EWs with errors for the detected lines in Figure \ref{EW_r_Z}. We do not find any clear correlation with distance or height. Two SiXIV detections with large EWs are at high latitudes and high heights. However, one high EW detection is for a small height and small latitude.  Thus this exercise indicates that most of the detected lines may not come from the diffuse ISM. 
\subsection{Comparison with simulations}
{The origin and location of this newly discovered super-virial phase were also investigated by different hydrodynamical simulations. 
\cite{Vijayan2022} showed that stellar feedback can produce this phase. However, their required star-formation rate was high compared to that of the Milky Way. \cite{Bisht2024} also claim that the super-virial hot phase arises from the heating of the virial hot gas by stellar feedback. Alternatively, \cite{Roy2024} showed the origin to be adiabatic compressive heating of the infalling virial hot gas. While the origin of the supervirial got gas phase in different simulations is different, they all point out that this phase is in the extra-planer region of the galaxy, which is in alignment with what we find in this paper. Observationally, we still can not say whether this gas is in the extra-planer region or in the extended CGM.}
\section{conclusion} \label{S: conclusion}
We aimed to explore the presence of super-virial gas at 
log(T/K) $\sim$ 7.5 using high-resolution X-ray spectra from Chandra's High-Energy Transmission Grating (HETG) observations of 27 XRBs (Table \ref{tab:source_description}). This study was motivated by the need to understand the distribution and origin of this hot gas phase, which remains debated since its unexpected discovery. By searching for signatures of this gas with SXVI K$\alpha$, SiXIV K$\alpha$, and NeX K$\alpha$ absorption lines, we sought to determine if this gas component is present in the diffuse ISM or is in the extraplanar region or in the extended CGM. Our analysis led to the following key findings:

\begin{itemize} 

\item \textbf{Limited Detection:} Highly ionized absorption lines were detected in only 7 out of the 30 XRB spectra (3 in \cite{Lara2024} and 27 in this paper), indicating that the super-virial gas is not abundantly present in the diffuse ISM (Figure \ref{lines} and Table \ref{fit}). Therefore, even if all the detected lines were assigned to the diffuse ISM, the covering fraction of the hot gas would be $<25\%$.
\item \textbf{Intrinsic Broad Features:} Two of the detected sources showed broad absorption lines, suggesting that these features are intrinsic to the XRB systems rather than originating from a diffuse ISM (Figure \ref{lines} and Table \ref{fit}). However, we can not conclude the same for the other 5 XRBs based on line width.  
\item \textbf{Line Variability:} All the detected lines, except for the SXVI line in the EXO 0748-676 sightline, exhibit temporal variability, reinforcing the hypothesis that these absorption features are tied to the dynamic environments around these variable XRB sources (Figure \ref{f:var}). For EXO 0748-676, non-coinciding data in 2003 suggests variability, but we can not have a definite conclusion due to a limited number of observations. 

\item \textbf{Lack of Correlation:} We found no significant correlation between the equivalent width of detected absorption lines and the distance, height, and latitude of the XRBs, again implying that the detected gas is not a part of a homogeneous diffuse ISM (Figure \ref{EW_r_Z}).
\end{itemize}

{From previous quasar absorption studies, it was difficult to infer the location of the super-virial gas; it could be anywhere from the ISM, to the extraplanar region, and/or in the extended CGM. With the critical test of absorption studies of Galactic XRBs in this paper,  we rule out ISM as a possible location of this gas phase.} Our work suggests that the super-virial temperature gas 
must reside in regions beyond the ISM. This may be an extra-planar region above the ISM or the extended outer CGM.  

\vspace{0.3cm}
\normalsize
\bibliography{main}
\bibliographystyle{aasjournal}




\newpage
\onecolumngrid
\begin{ThreePartTable}
\begin{TableNotes}
\end{TableNotes}
\begin{longtable}{|c|c|c|c|c|c|c|}
    \caption{Galactic Source Description} \label{tab:source_description}\\ 
    \hline
    \textbf{Source} & \textbf{Galactic Latitude (degree)} & \textbf{Distance (kpc)} & \textbf{Height (kpc)} & \textbf{Obs. ID} & \textbf{Exp. Time (ks)} & \textbf{Exp. Mode} \\ 
    \hline
    \endfirsthead
    \caption*{(Continued)}\\
    \hline
    \textbf{Source} & \textbf{Galactic Latitude (degree)} & \textbf{Distance (kpc)} & \textbf{Height (kpc)} & \textbf{Obs. ID} & \textbf{Exp. Time (ks)} & \textbf{Exp. Mode} \\
    \hline
    \endhead
    \hline
    \multicolumn{6}{|r|}{\textit{Continued on next page}} 
    \endfoot
    \hline
    \endlastfoot

        4U1636-53 & -4.818 & 6.0 & 0.5  & 6635 & 23.07 & CC \\
        & & & & 105 &  29.40 &  CC \\
         &  &  &  & 1939 & 26.29 & TE \\
        & & & & 22701 &  46.61 &  TE \\
        &  &  &  & 22936 & 30.13 & TE \\
        & & & & 24625 &  29.09 &  TE \\
        &  &  &  & 24626 & 24.40 & TE \\
        & & & & 20791 &  62.93 &  TE \\
        &  &  &  & 21099 & 31.99 & TE \\
        & & & & 21100 &  32.95 &  TE \\
        \hline
        EXO 0748-676 & -19.811 & 8.0; 5.7 & 2.88,2.05  & 1017 & 47.99 & TE \\
         &  & &  & 4573 & 162.89 & TE \\
        &  &  &  & 4574 & 123.54 & TE \\
        \hline
        PSRB 0833-45 & -2.787 & 0.3 & 0.015  & 131 & 35.95 & CC \\
        \hline
        SAX J1808.4-3658 & -8.148 & 2.8 & 0.4  & 6297 & 14.29 & CC \\
        \hline
        Swift J1753.5-0127 & +12.186 & -- & --  & 14428 & 20.02 & CC \\
        \hline
        Swift J1910.2-0546 & -6.844 & -- & --  & 14634 & 29.96 & CC \\
        \hline
        4U 1728-16 (GX9+9) & +9.038 & 4.4;7.5 & 0.69,1.19  & 703 & 20.72 & TE \\
        & &  &  & 11072 & 95.84 & TE \\
        \hline
        V*V 821Ara (GX 339-4) & -4.326 & 10.0 & 0.76  & 4569 & 49.90 & CC \\
         &  &  &  & 4570 & 38.90 & CC \\
         &  &  & & 4571 & 35.32 & CC \\
          &  &  & & 4420 & 74.05 & TE \\
          &  &  & & 5475 & 37.39 & TE \\
          &  &  & & 6290 & 21.22 & TE \\
        \hline
        GS 1826-238 & -6.088 & 7.5 & 0.8  & 2739 & 68.23 & TE \\
        \hline
        4U 2129+12 & -27.312 & 5.8 & 2.99  & 675 & 19.80 & TE \\
         &  &  &   & 4572 & 59.17 & TE \\
         &  &  &   & 27459 & 9.33 & TE \\
        \hline
        4U 1543-624 & -6.337 & 7.0 & 0.77  & 702 & 27.40 & TE \\
        \hline
        XTE J1650-500 & -3.427 & -- & -- & 2699 & 19.21 & CC \\
        &  &  &   & 2700 & 28.50 & CC \\
        & & & & 3400 & 10.00 & TE \\
        & & & & 3401 & 9.51 & TE \\
        \hline
        4U 1254-69 & -6.4 & 13.0$\pm3.0$ & 1.45$\pm0.3$ & 3823 & 51.69 & TE \\
        \hline
         4U 1957+11 & -9.33 & -- & -- & 10660 & 18.98 & CC \\
        &  &  &   & 4552 & 65.60 & TE \\
        & & & & 10659 & 9.87 & TE \\
        & & & & 10661 & 9.82 & TE \\
        \hline
        4U 0614+091 & -3.36 & 2.2$\pm0.7$ & 0.13$\pm0.04$ & 10759 & 58.88 & TE \\
        &  &  &   & 10760 & 44.10 & TE \\
        & & & & 10857 & 57.28 & TE \\
        & & & & 10858 & 34.36 & TE \\
        \hline
        4U 1705-44 & -2.34 & 7.6$\pm0.3$ & 0.3$\pm0.01$ & \textcolor{black}{1924} & \textcolor{black}{5.86} & \textcolor{black}{CC} \\
        &  &  &   & 1923 & 24.41 & TE \\
        & & & & 5500 & 26.47 & TE \\
        & & & & 18086 & 23.95 & TE \\
        & & & & 19451 & 37.98 & TE \\
        & & & & 20082 & 69.40 & TE \\
        \hline
        4U 1728-34 & -0.15 & 5.2$\pm0.5$ & 0.014$\pm0.001$
                  & 6567 & 151.25 & CC \\
        &  &  &   & 6568 & 49.30 & CC \\
        &  &  & & 7371 & 39.56 & CC \\
        & & & & 2748 & 29.62 & TE \\
        & & & & 19452 & 20.23 & TE \\
    
         & & & & 20106 & 58.42 & TE \\
          & & & & 20107 & 32.43 & TE \\
        \hline
        4U 1735-44 & -6.99 & 9.4$\pm1.4$ & 1.15$\pm0.17$
                  & 6637 & 24.09 & CC \\
        &  &  &   & 6638 & 23.03 & CC \\
        & & & & 704 & 24.35 & TE \\
        & & & & 25248 & 26.18 & TE \\
        & & & & 25897 & 14.41 & TE \\
        & & & & 25898 & 10.71 & TE \\
        & & & & 25899 & 33.82 & TE \\
        & & & & 25900 & 14.41 & TE \\
        & & & & 25901 & 24.06 & TE \\
        & & & & 27532 & 17.00 & TE \\
        & & & & 27917 & 24.25 & TE \\
        & & & & 28919 & 22.32 & TE \\
        \hline
        4U 1820-30 & -7.91 & 7.6$\pm0.4$ & 1.05$\pm0.05$
        &  6633 & 25.08 & CC \\
        &  &  &   & 6634 & 25.03 & CC \\
        &  &  &   & 22276 & 13.65 & CC \\
        &  &  &   & 22277 & 14.62 & CC \\ 
        &  &  &   & 24698 & 25.54 & CC \\
       
        &  &  &   & 25029 & 25.86 & CC \\
        &  &  &   & 25030 & 23.49 & CC \\
        &  &  &   & 25031 & 22.99 & CC \\
        &  &  &   & 25032 & 20.04 & CC \\
        &  &  &   & 25033 & 23.38 & CC \\
        &  &  &   & 25037 & 17.81 & CC \\
        &  &  &   & 25038 & 15.50 & CC \\
        &  &  &   & 7032 & 46.16 & CC \\
        & & & & 1021 & 9.64 & TE \\
        & & & & 1022 & 10.65 & TE \\
        \hline
        4U 1626-67 & -13.09 & $3.5^{+0.2}_{-0.3}$ & 0.8$^{+0.1}_{-0.1}$
        & 104 & 39.48 & TE \\
        &  &  &   & 3504 & 94.81 & TE \\
        & & & & 11058 & 76.87 & TE \\
        & & & & 17448 & 48.90 & TE \\
        & & & & 21686 & 45.88 & TE \\
        & & & & 24700 & 11.76 & TE \\
        & & & & 24768 & 17.13 & TE \\
        & & & & 26204 & 17.13 & TE \\
        & & & & 26250 & 12.73 & TE \\
        & & & & 26009 & 56.81 & TE \\
        & & & & 26086 & 29.34 & TE \\
        & & & & 27954 & 29.34 & TE \\
        & & & & 27960 & 16.16 & TE \\
        & & & & 28366 & 14.50 & TE \\
        & & & & 19782 & 19.85 & TE \\
        & & & & 20909 & 25.14 & TE \\
        \hline
        Serpens X-1 & 4.84 & 11.1$\pm1.6$ & 0.94$\pm0.13$
        & 16208 & 142.04 & CC \\
        & & & & 16209 & 156.54 & CC \\
        & & & &700 & 76.42 & TE \\
        & & & & 17485 & 83.26 & TE \\
        & & & & 17600 & 37.05 & TE \\
        \hline
        Cygnus X-1 & -11.31 & 13.4$\pm2.0$ & 2.7$\pm0.4$
        & 1511 & 12.62 & CC \\
        &  &  &   & 2415 & 30.02 & CC \\
        &&&& 3407 & 16.07 & CC \\
        &&&& 3724 & 8.10 & CC \\
        &&&& 3815 & 55.63 & CC \\
        &&&& 12314 & 0.90 & CC \\
        &&&& 12472 & 3.31 & CC \\
        &&&& 107 & 11.40 & TE \\
        &&&& 2741 & 1.89 & TE \\ 
        &&&& 2742 & 1.87 & TE \\ 
        \hline 
        & & & & 2743 & 2.42 & TE \\
        & & & & 3814 & 47.19 & TE \\  
        &&&& 8525 & 29.43 & TE \\
        &&&& 9847 & 18.86 & TE \\
        &&&& 11044 & 29.43 & TE \\
        &&&& 12313 & 2.14 & TE \\
        &&&& 13219 & 4.42 & TE \\
        &&&& 16735 & 17.57 & TE \\
        &&&& 26685 & 7.19 & TE \\
        &&&& 29064 & 9.06 & TE \\ 
        \hline 
        GX 349+2 & 2.74 & 9.2 & 0.44
        & 6628 & 12.54 & CC \\
        &  &  &   & 7336 & 12.06 & CC \\
        &&&& 12199 & 18.77 & CC \\
        &&&& 13221 & 16.52 & CC \\
        &&&& 13222 & 25.81 & CC \\
        &&&& 715 & 9.42 & TE \\
        &&&& 14256 & 9.30 & TE \\
        &&&& 3354 & 17.45 & TE \\
        &&&& 13220 & 18.87 & TE \\
        &&&& 18084 & 20.42 & TE \\
        \hline
        GX 340+00 & -0.07 & 11.0$\pm0.3$ & 0.0130$\pm0.0004$
        & 1922 & 5.82 & CC \\
        &  &  &   & 6631 & 25.03 & CC \\
        &&&& 6632 & 23.57 & CC \\
        &&&& 1921 & 23.36 & TE \\
        &&&& 18085 & 24.11 & TE \\
        &&&& 19450 & 60.15 & TE \\
        &&&& 20099 & 60.62 & TE \\
        \hline
        GX 5-1 & -1.02 & 9.2 & 0.16
        & 10691 & 8.05 & CC \\
        &  &  &   & 10692 & 8.15 & CC \\
        &&&& 10693 & 8.11 & CC \\
        &&&& 10694 & 3.75 & CC \\
        &&&& 5888 & 44.98 & CC \\
        &&&& 716 & 8.91 & TE \\
        &&&& 22411 & 15.67 & TE \\
        &&&& 22412 & 17.56 & TE \\
        &&&& 22413 & 16.25 & TE \\
        &&&& 22414 & 17.28 & TE \\
        &&&& 22415 & 15.76 & TE \\
        &&&& 22416 & 14.00 & TE \\
        &&&& 19449 & 80.49 & TE \\
        &&&& 20119 & 12.10 & TE \\
        \hline
        GX 3+1 & 0.79 & $5.0^{+0.8}_{-0.7}$ & 0.07$^{+0.01}_{-0.009}$ 
        & 2745 & 8.96 & CC \\
        &  &  &   & 24701 & 20.38 & TE\\
        &&&& 24769 & 27.15 & TE \\
        &&&& 24770 & 17.48 & TE \\
        &&&& 26432 & 21.06 & TE \\
        &&&& 26434 & 11.68 & TE \\ 
        &&&& 27271 & 10.25 & TE \\
        &&&& 27272 & 9.79 & TE \\
        &&&& 27273 & 9.79 & TE \\
        &&&& 16307 & 43.59	& TE \\
        &&&& 16492 & 43.59 & TE \\
        &&&& 18615 & 12.16	& TE \\
        &&&& 19890 & 29.08 & TE \\
        &&&& 19907 & 26.01 & TE \\
        &&&& 19957 & 29.08	& TE \\
        &&&& 19958 & 29.08	& TE \\
        \hline
        GX 13+1 & 0.10 & 7$\pm1$ & 0.01$\pm0.002$
        & 11818	 & 24.41 & CC \\
        &  &  &   & 13197 & 10.09 & CC \\ 
        &&&& 2708 & 29.35 & TE \\
        &&&& 11814 & 28.12	 & TE \\
        &&&& 11815	& 28.12	 & TE \\
        &&&& 11816 & 28.12	& TE \\
        &&&& 11817 & 28.12	& TE \\
        &&&& 20191 & 24.25	 & TE \\
        &&&& 20192	& 22.32	 & TE \\
        &&&& 20193	& 24.25 & TE \\
        &&&& 20194 & 24.56 & TE \\
        \hline
\end{longtable}
\end{ThreePartTable}

\begin{ThreePartTable}
\begin{TableNotes}
\item[*] $1\sigma$ errors are given otherwise $1\sigma$ upper-limits are quoted. Numbers are made bold whenever lines are detected with $3\sigma$.
\item[**] fitted with a broad Gaussian 
\end{TableNotes}
\begin{longtable}{|c|c|c|c|c|c|} 
    \caption{Best-fit parameter value}\label{fit}\\
    \hline
    &  &  &\multicolumn{3}{c|}{\textbf{Equivalent Width (mA)\tnote{*}}}\\
    \cline{4-6}
    \textbf{Source} & \textbf{Exp. Mode} & \textbf{Grating} &Ne X & Si XIV & S XVI \\
    & & & (12.132 A) & (6.181 A) & (4.728 A) \\
    \hline
    \endfirsthead
    \caption*{(Continued)}\\
    \hline
    &  &  &\multicolumn{3}{c|}{\textbf{Equivalent Width (mA)\tnote{*}}}\\
    \cline{4-6}
    \textbf{Source} & \textbf{Exp. Mode} & \textbf{Grating} &Ne X & Si XIV & S XVI \\
    & & & (12.132 A) & (6.181 A) & (4.728 A) \\
    \hline
    \endhead
    \hline
    \multicolumn{6}{|r|}{\textit{Continued on next page}} \\
    \hline
    \endfoot
    \hline
    \insertTableNotes\\
    \endlastfoot
            4U1636-53 & CC & HEG & -2.11  & -0.63 & -0.32 \\
             &  & MEG & -1.26 & -0.28 & -0.19 \\
             & TE & HEG & -1.92  & -0.06 & -0.4\\
             &  & MEG & -0.44 & \textbf{-1.04$^{+0.2}_{-0.2}$} & -0.33$^{+0.28}_{-0.28}$  \\
            \hline
            EXO0748-676 & TE & HEG & -5.17$^{+4.14}_{-4.14}$  & \textbf{-5.11$^{+0.7}_{-0.7}$} & \textbf{-3.19$^{+0.95}_{-0.95}$} \\
             &  & MEG & -4.34 & \textbf{-6.43$^{+0.68}_{-0.68}$} & -1.52$^{+1.0}_{-1.0}$ \\
            \hline
            PSRB0833-45 & CC & HEG & -4.55  & {-4.28$^{+2.68}_{-2.68}$} & -3.89 \\
             &  & MEG & -5.15 & -4.46 & -7.75 \\
            \hline
            SAXJ1808.4-3658 & CC & HEG & -2.21$^{+2.21}_{-2.21}$  & -0.35 & -3.01 \\
             &  & MEG & -1.7 & -0.2 & -2.59 \\
            \hline
            SwiftJ1753.5-0127 & CC & HEG & -0.46  & 0.83$^{+0.63}_{-0.63}$ & -1.99$^{+1.16}_{-1.16}$ \\
             &  & MEG & -0.97 & -0.15 & -1.92 \\
            \hline
            SwiftJ1910.2-0546 & CC & HEG & -0.66  & -0.11 & -0.29 \\
             &  & MEG & -0.52$^{+0.41}_{-0.41}$ & {-0.43$^{+0.24}_{-0.24}$} & -0.62 \\
            \hline
            4U1728-16 (GX9+9) & TE & HEG & -0.74  & -0.17 & -0.26$^{+0.23}_{-0.23}$ \\
             &  & MEG & -0.51$^{+0.46}_{-0.46}$ & \textbf{-0.73$^{+0.18}_{-0.18}$} & -0.42 \\
             \hline
             V*V821Ara (GX 339-4) & CC & HEG & -0.56$^{+0.31}_{-0.31}$  & -0.02 & -0.17 \\
             &  & MEG & -0.2 & -0.01 & -0.03 \\
             & TE & HEG & -1.46  & -0.27 & -0.14 \\
             &  & MEG & -0.34 & \textbf{-0.84$^{+0.23}_{-0.23}$} & -0.22 \\
            \hline
            GS1826-238 & TE & HEG & -7.94  & -0.5 & -0.83 \\
             &  & MEG & -1.54 & -0.59$^{+0.55}_{-0.55}$ & -1.34 \\
            \hline
            4U2129+12 & TE & HEG & -2.09  & -0.24 & -0.63 \\
             &  & MEG & -0.65 & -0.15 & -0.70 \\
            \hline
            4U1543-624 & TE & HEG & -3.43  & -1.73 & -2.02 \\
             &  & MEG & -0.96 & -0.33 & -0.5 \\
            \hline
            XTEJ1650-500 & CC & HEG & -0.03 (c)  & -0.16$^{+0.15}_{-0.15}$ & -0.06 \\
             &  & MEG & 0.22 & -0.13$^{+0.12}_{-0.12}$ & -0.17 \\
             XTEJ1650-500 & TE & HEG & NE  & NE & NE \\
             &  & MEG & NE & NE & NE \\
             \hline
             \hline
        4U 1254-69 & TE & HEG & -3.85  & -1.23 & -2.17$^{+1.11}_{-1.11}$ \\
             &  & MEG & -1.63 & -1.1$^{+0.58}_{-0.58}$ & -0.77 \\
             \hline
         4U 1957+11 & CC & HEG & -6.29  & -0.6 & -0.79 \\
             &  & MEG & -2.08 & -0.96$^{+0.81}_{-0.81}$ & -0.54 \\
             & TE & HEG & -1.13 & -0.74 & -0.76$^{+0.51}_{-0.51}$ \\
             &  & MEG & -0.26 & -0.69 & -0.66 \\
             \hline
        4U 0614+091 
            & TE & HEG & -0.43 & -0.36 & -0.51 \\
             &  & MEG & -0.13 & -0.58$^{+0.23}_{-0.23}$  & -0.16 \\
            \hline
        4U 1705-44 & CC & HEG & -35.05  & -0.62 & -0.36 \\
             &  & MEG & -13.44$^{+10.24}_{-11.54}$ & -0.21 & -0.40 \\
             & TE & HEG & -18.6  & -0.11 & -0.1 \\
             &  & MEG & -2.17 & -0.1 & -0.39$^{+0.28}_{-0.28}$  \\
            \hline
        4U 1728-34 & CC & HEG & -3.8  & -0.37 & -0.31 \\
             &  & MEG & -1.96 & -0.25$^{+0.21}_{-0.21}$ & -0.66$^{+0.23}_{-0.23}$ \\
             & TE & HEG & -736.24  & -0.74 & -0.36 \\
             &  & MEG & -131.7 & -0.43 & -0.69 \\
            \hline
            4U 1735-44 & CC & HEG & -2.05  & -0.07 & -0.48$^{+0.43}_{-0.43}$ \\
             &  & MEG & -0.89$^{+0.81}_{-0.81}$ & -0.31 & -0.68 \\
             & TE & HEG & -2.11  & -0.31 & -0.13 \\
             &  & MEG & -1.06$^{+1.05}_{-1.05}$ & -0.59$^{+0.44}_{-0.44}$ & -0.87$^{+0.61}_{-0.61}$ \\
            \hline
             4U 1820-30 & CC & HEG & 1.09$^{+0.41}_{-0.41}$   & -0.05 & -0.36 \\
             &  & MEG & -0.07 & -0.14 & -0.4 \\
             & TE & HEG & -1.41$^{+1.11}_{-1.11}$  & -0.42 & -0.57 \\
             &  & MEG & -0.33 & \textbf{-1.61$^{+0.42}_{-0.42}$} & -0.39 \\
            \hline
            4U 1626-67 & TE & HEG & -0.21 (-2.01)    & -0.46 & -0.65 \\
             &  & MEG & -0.06 (-0.91$^{+0.87}_{-0.87}$) & -0.34 & -0.22 \\
            \hline
           Serpens X-1 & CC & HEG & -1.13 & -0.02 & -0.03 \\
             &  & MEG & -0.28 &  -0.01 & -0.07 \\
             & TE & HEG & 1.18  & -0.05 & -0.07 \\
             &  & MEG & -0.29 & -0.18$^{+0.16}_{-0.16}$ & -0.06 \\
            \hline
            Cygnus X-1 & CC & HEG & \textbf{-2.52$^{+0.2}_{-0.19}$ $^{**}$}  & \textbf{-4.59$^{+0.15}_{-0.13}$ $^{**}$} & \textbf{-2.69$^{+0.24}_{-0.23}$ $^{**}$} \\
             &  & MEG & \textbf{-1.74$^{+0.36}_{-0.22}$ $^{**}$} & \textbf{-6.07$^{+0.12}_{-0.12}$ $^{**}$}  & \textbf{-2.85$^{+0.34}_{-0.21}$ $^{**}$} \\
             & TE & HEG & -0.31  & \textbf{-5.47$^{+0.19}_{-0.46}$  $^{**}$ } & \textbf{-3.19$^{+0.34}_{-0.34}$ $^{**}$ }\\
             &  & MEG & -1.07 & \textbf{-4.63$^{+0.25}_{-0.25}$ $^{**}$}  & \textbf{-2.21$^{+0.25}_{-0.29}$ $^{**}$ } \\
            \hline
            GX 349+2 & CC & HEG & -1.27 & -0.14  & -0.18  \\
             &  & MEG & -0.79 &  -0.13 & -0.11 \\
             & TE & HEG & -1.2  & -0.17 & -0.22$^{+0.2}_{-0.2}$ \\
             &  & MEG & -0.98 & -0.6 & -0.15 \\
            \hline
            GX 340+00 & CC & HEG & -18.74 & -0.89 & -0.49 \\
             &  & MEG & -11.27 &  -0.32 & -0.12 \\
             & TE & HEG & -1806  & -0.26 & -0.32  \\
             &  & MEG & -640 & -0.09  & -0.21 \\
            \hline
            GX 5-1 & CC & HEG & -13.75 & -0.12 & -0.42$^{+0.2}_{-0.2}$ \\
             &  & MEG & -11.22 &  -0.23 & -0.37 \\
             & TE & HEG & -358.16  & -0.19 & -0.10  \\
             &  & MEG & -89.94 & -0.15  & -0.25$^{+0.13}_{-0.13}$ \\
            \hline
             GX 3+1 & CC & HEG & -24.2 & -0.19 & -0.57 \\
             &  & MEG & -12.97$^{+8.38}_{-9.33}$ &  -1.15 & -0.33 \\
             & TE & HEG & -1134.6 & -0.13 & -0.17 \\
             &  & MEG & -3.91 &  -0.25 & -0.11 \\
            \hline
            GX 13+1 & CC & HEG & -19.15 & \textbf{-6.38$^{+0.75}_{-0.75}$ $^{**}$} & \textbf{-2.8$^{+0.5}_{-0.38}$}  \\
             &  & MEG & -19.14 & \textbf{-5.47$^{+0.58}_{-0.44}$ $^{**}$} & \textbf{-2.99$^{+0.41}_{-0.41}$}  \\
             & TE & HEG & -2968.09 & \textbf{-7.15$^{+0.3}_{-0.27}$ $^{**}$}  & \textbf{-4.75$^{+0.27}_{-0.26}$ $^{**}$}\\
             &  & MEG & -11.98$^{+10.47}_{-8.52}$ & \textbf{-7.01$^{+0.17}_{-0.21}$ $^{**}$} & \textbf{-4.52$^{+0.28}_{-0.12}$ $^{**}$} \\
            \hline 
            
\end{longtable}
\end{ThreePartTable}

\begin{ThreePartTable}
\begin{TableNotes}
\item[*] $1\sigma$ errors are given otherwise $1\sigma$ upper-limits are quoted. 
\end{TableNotes}
\begin{longtable}{|c|c|c|c|c|c|c|}
    \caption{Variability test}\label{variability}\\
    \hline
    \textbf{Source} & \textbf{Exp. Mode} & \textbf{Grating} & \textbf{Ion} & \textbf{ObsId (Year of Obs)} & \textbf{Equivalent Width (mA)\tnote{*}} & \textbf{Variable} \\ 
    \hline
    \endfirsthead
    \caption*{(Continued)}\\
    \hline
    \textbf{Source} & \textbf{Exp. Mode} & \textbf{Grating} & \textbf{Ion} & \textbf{ObsId (Year of Obs)} & \textbf{Equivalent Width (mA)\tnote{*}}& \textbf{Variable} \\ 
    \hline
    \endhead
    \hline
    \multicolumn{7}{|r|}{\textit{Continued on next page}} \\
    \hline
    \endfoot
    \hline
    \insertTableNotes\\
    \endlastfoot
    
    4U1636-53 & TE & MEG & SiXIV & Total (XX) & {-1.04$^{+0.2}_{-0.2}$}  & Yes\\ 
    & & & & 105 (1999) & {-1.77$^{+0.4}_{-0.4}$} & \\
    & & & & 1939 (2001) & -1.01$^{+0.37}_{-0.37}$ & \\
    & & & & 20791 (2018) & -0.86$^{+0.54}_{-0.54}$ & \\
    & & & & 21099 (2018) & -1.41 &  \\
    & & & & 21100 (2018) & -0.91 &  \\
    & & & & 22701 (2021) & -0.73 & \\
    & & & & 22936 (2021) & -1.09$^{+0.97}_{-0.97}$ & \\
    & & & & 24625 (2021) & -1.89 & \\
    & & & & 24626 (2021) & -2.1 & \\
    \hline
    4U1728-16 (GX9+9) & TE & MEG & SiXIV & Total (XX) & {-0.73$^{+0.18}_{-0.18}$} & Yes \\ 
    & & & & 703 (2000) & {-1.43$^{+0.4}_{-0.4}$} &\\
    & & & & 11072 (2010) & {-0.56$^{+0.2}_{-0.2}$} & \\
    \hline
    VstarV821Ara (GX 339-4) & TE & MEG & SiXIV & Total (XX) & {-0.84$^{+0.23}_{-0.23}$}  & Yes \\ 
    & & & & 4420 (2003) & -0.32$^{+0.23}_{-0.23}$ & \\
    & & & & 5475 (2005) & -3.92$^{+2.55}_{-2.55}$ &\\
    & & & & 6290 (2005) & -20.6 &\\
    \hline
    EXO0748-676 & TE & HEG & SiXIV & Total (XX) & {-5.11$^{+0.7}_{-0.7}$}  & Yes \\ 
    & & & & 1017 (2001) & -4.56 &\\
    & & & & 4573 (2003) & {-5.53$^{+0.97}_{-0.97}$} & \\
    & & & & 4574 (2003) & {-5.19$^{+1.3}_{-1.3}$} &\\
    & & & SXVI & Total (XX) & {-3.19$^{+0.95}_{-0.95}$} & No \\ 
    & & & & 1017 (2001) & -7.99 &\\
    & & & & 4573 (2003) & -3.14$^{+1.46}_{-1.46}$ &\\
    & & & & 4574 (2003) & -2.2$^{+1.81}_{-1.81}$ &\\
    & & MEG & SiXIV & Total (XX) & {-6.43$^{+0.68}_{-0.68}$}  & Yes\\ 
    & & & & 1017 (2001) & -6.16 &\\
    & & & & 4573 (2003) & {-7.28$^{+0.9}_{-0.9}$} &\\
    & & & & 4574 (2003) & {-5.73$^{+1.19}_{-1.19}$} &\\
    \hline
    4U 1820-30 & TE & MEG & SiXIV & Total (XX) & {-1.61$^{+0.42}_{-0.42}$} & Yes\\ 
    & & & & 1021 (2001) & -1.12 &\\
    & & & & 1022 (2001) & {-2.58$^{+0.57}_{-0.57}$} &\\
    \hline
    GX 13+1 & CC & HEG & SiXIV & Total (XX) & {-6.38$^{+0.75}_{-0.75}$ }  & Yes\\
    & & & & 11818 (2010) & {-4.52$^{+0.55}_{-1.01}$} &\\ 
    & & & & 13197 (2011) & -9.41$^{+1.34}_{-1.48}$ &\\
    & & MEG & SiXIV & Total (XX) & {-5.47$^{+0.58}_{-0.44}$ } &\\
    & & & & 11818 (2010) & {-4.45$^{+0.55}_{-0.73}$} &\\
    & & & & 13197 (2011) & -7.52$^{+0.65}_{-1.03}$ &\\
    & TE & HEG & SiXIV & Total (XX) & {-7.15$^{+0.3}_{-0.27}$ } &\\
    & & & & 2708 (2002) & -7.61$^{+0.8}_{-0.71}$ & \\
    & & & & 11814 (2010) & 6.54$^{+1.01}_{-1.04}$ &\\
    & & & & 11815 (2010) & -7.59$^{+0.79}_{-0.84}$ &\\
    & & & & 11816 (2010) & -3.98$^{+0.8}_{-0.76}$ &\\
    & & & & 11817 (2010) & -7.3$^{+0.45}_{-0.36}$ &\\
    & & & & 20191 (2018) & -7.28$^{+1.01}_{-1.12}$ &\\
    & & & & 20192 (2018) & -6.62$^{+0.78}_{-0.81}$ &\\
    & & & & 20193 (2019) & -7.6$^{+1.06}_{-1.15}$ &\\
    & & & & 20194 (2019) & -13.7$^{+0.93}_{-1.01}$  &\\
    && MEG & SiXIV & Total (XX) &  {-7.01$^{+0.17}_{-0.21}$ } &\\
    & & & & 2708 (2002) & -7.67$^{+0.36}_{-0.72}$ &\\
    & & & & 11814 (2010) & 6.31$^{+0.86}_{-0.36}$ &\\
    & & & & 11815 (2010) & -7.66$^{+0.68}_{-0.71}$ &\\
    & & & & 11816 (2010) & -3.27$^{+0.47}_{-0.5}$ &\\
    & & & & 11817 (2010) & -6.74$^{+0.85}_{-0.52}$ &\\
    & & & & 20191 (2018) & -6.31$^{+0.42}_{-0.9}$ &\\
    & & & & 20192 (2018) & -7.89$^{+0.81}_{-0.56}$ &\\
    & & & & 20193 (2019)& -6.08$^{+0.7}_{-0.75}$ &\\
    & & & & 20194 (2019)& -12.56$^{+0.72}_{-1.00}$ &\\
    & CC & HEG & SXVI & Total (XX) & {-2.8$^{+0.5}_{-0.38}$} &\\
    & & & & 11818 (2010) & -2.17$^{+0.49}_{-0.72}$ &\\
    & & & & 13197 (2011) & -3.73$^{+0.76}_{-0.69}$ &\\
    & & MEG & SXVI & Total (XX) & {-2.99$^{+0.41}_{-0.41}$}&\\
    & & & & 11818 (2010) & -3.56$^{+1.63}_{-0.59}$ &\\
    & & & & 13197 (2011) & -3.99$^{+0.71}_{-0.69}$  &\\
    & TE & HEG & SXVI & Total (XX) & {-4.75$^{+0.27}_{-0.26}$ } &\\
    & & & & 2708 (2002)& -5.44$^{+0.6}_{-0.79}$ &\\
    & & & & 11814 (2010) & -2.25$^{+0.7}_{-0.96}$ &\\
    & & & & 11815 (2010) & -4.89$^{+1.03}_{-1.03}$  &\\
    & & & & 11816 (2010) & -2.15$^{+0.75}_{-1.19}$ &\\
    & & & & 11817 (2010) & -3.05$^{+0.53}_{-0.6}$ &\\
    & & & & 20191 (2018)& -3.93$^{+0.75}_{-0.67}$ &\\
    & & & & 20192 (2018)& -7.0$^{+0.654}_{-1.0}$ &\\
    & & & & 20193 (2019)& -5.25$^{+0.49}_{-1.18}$ &\\
    & & & & 20194 (2019)& -6.43$^{+0.74}_{-0.17}$ &\\
    && MEG & SXVI & Total (XX) & {-4.52$^{+0.28}_{-0.12}$ } &\\
    & & & & 2708 (2002)& -4.54$^{+0.42}_{-0.71}$ &\\
    & & & & 11814 (2010)& -3.42$^{+0.97}_{-0.7}$ &\\
    & & & & 11815 (2010)& -4.69$^{+0.71}_{-0.67}$ &\\
    & & & & 11816 (2010)& -2.54$^{+0.81}_{-0.18}$ &\\
    & & & & 11817 (2010)& -3.89$^{+0.89}_{-0.47}$ &\\
    & & & & 20191 (2018)& -4.52$^{+0.57}_{-0.51}$ &\\
    & & & & 20192 (2018)& -4.86$^{+0.38}_{-0.9}$ &\\
    & & & & 20193 (2019)& -3.89$^{+0.54}_{-0.88}$ &\\
    & & & & 20194 (2019)& -9.90$^{+0.6}_{-0.5}$ &\\
    \hline
    Cygnus X-1 &  CC & HEG & Ne X & Total (XX) & {-2.52$^{+0.2}_{-0.19}$ } & Yes\\
        & &&& 1511 (2000) & -0.69$^{+0.23}_{-2.42}$ &\\
        &  &  &   & 2415 (2001) & -2.34$^{+0.33}_{-1.2}$ &\\
        &&&& 3407 (2001) & -1.0$^{+0.4}_{-0.3}$ &\\
        &&&& 3724 (2002)& -5.87$^{+0.31}_{-0.54}$ & \\
        &&&& 3815 (2003)& -1.0$^{+0.3}_{-0.4}$&\\
        &&&& 12314 (2010)& -1.56 &\\
        &&&& 12472 (2011)& -12.76 &\\
        &   & MEG & Ne X & Total (XX) & {-1.74$^{+0.36}_{-0.22}$ } & \\
        & &&& 1511 (2000) & -4.32$^{+1.69}_{-0.32}$ &\\
        &  &  &   & 2415  (2001) & -2.27$^{+0.49}_{-0.56}$ &\\
        &&&& 3407  (2001)& -0.78&\\
        &&&& 3724 (2002) & -4.75$^{+0.45}_{-0.72}$&\\
        &&&& 3815 (2003) & -2.39$^{+0.17}_{-0.77}$&\\
        &&&& 12314 (2010) & -22.23 &\\
        &&&& 12472 (2011) & -739.59 &\\
&  CC & HEG & Si XIV & Total (XX) & {-4.59$^{+0.15}_{-0.13}$ } &\\
        & &&& 1511 (2000) & -5.46$^{+0.92}_{-0.2}$&\\
        &  &  &   & 2415 (2001) & -3.78$^{+0.12}_{-0.65}$ &\\
        &&&& 3407  (2001) & -2.33$^{+0.25}_{-0.33}$&\\
        &&&& 3724 (2002) & -11.03$^{+0.66}_{-0.04}$ &\\
        &&&& 3815 (2003) & -5.35$^{+0.2}_{-0.19}$ &\\
        &&&& 12314 (2010) & -1.05$^{+0.7}_{-0.7}$ &\\
        &&&& 12472 (2011) & -2.44 &\\
        &   & MEG & Si XIV & Total (XX) & {-6.07$^{+0.12}_{-0.12}$ } &\\
        & &&& 1511 (2000) & -5.96$^{+0.49}_{-0.82}$ &\\
        &  &  &   & 2415 (2001) & -5.03$^{+0.56}_{-0.21}$&\\
        &&&& 3407 (2001) & -3.34$^{+0.29}_{-0.24}$ &\\
        &&&& 3724 (2002) & 12.24$^{+0.75}_{-0.014}$ &\\
        &&&& 3815 (2003) & -5.68$^{+0.36}_{-0.02}$ &\\
        &&&& 12314 (2010) & -0.16 &\\
        &&&& 12472 (2011) & -1.72$^{+0.39}_{-0.76}$ &\\
        
        &TE&HEG&Si XIV&Total (XX) & {-5.47$^{+0.19}_{-0.46}$   } &\\ 
        &&&&107 (1999) & -30.64$^{+12.02}_{-21.63}$ &\\
        &&&& 2741 (2002) & -44.67$^{+11.39}_{-20.46}$ & \\ 
        &&&& 2742 (2002) & -46.75$^{+11.51}_{-14.55}$ &\\ 
        & & & & 2743 (2002) & -161.89$^{+9.5}_{-11.18}$ & \\
        & & & & 3814 (2003) & -9.43$^{+0.72}_{-0.37}$ &\\  
        &&&& 8525 (2008) & -10.89$^{+0.80}_{-0.41}$  &\\
        &&&& 9847 (2008) & -8.62$^{+0.60}_{-1.09}$ &\\
        &&&& 11044 (2010)& -1750.88 &\\
        &&&& 12313 (2010) &  -95.24$^{+54.83}_{-42.55}$ &\\
        &&&& 13219 (2011)& -86.07$^{+47.48}_{-42.48}$ &\\
        &&&& 16735 (2016)& -3.29$^{+0.65}_{-0.02}$ &\\
        &&MEG&Si XIV& Total (XX) & {-4.63$^{+0.25}_{-0.25}$ } & \\ 
        &&&&107 (1999) & -416.89$^{+12.19}_{-364.02}$ &\\
        &&&& 2741 (2002) & -683.04$^{+23.01}_{-1316.14}$ & \\ 
        &&&& 2742 (2002) & -4.31$^{+2.29}_{-5.04}$ &\\ 
        & & & & 2743 (2002) & -742.67$^{+29.5}_{-1659.5}$ &  \\
        & & & & 3814 (2003)& -10.30$^{+1.02}_{-0.29}$ &\\  
        &&&& 8525 (2008)& -9.48$^{+0.38}_{-1.04}$  &\\
        &&&& 9847 (2008)& -7.61$^{+0.57}_{-1.14}$ &\\
        &&&& 11044 (2010)&  -3.84 &\\
        &&&& 12313 (2010)& -733.76$^{+17.14}_{-471.24}$ &\\
        &&&& 13219 (2011)& -757.86$^{+13.30}_{-986.21}$ &\\
        &&&& 16735 (2016)& -6.43$^{+1.71}_{-1.71}$ &\\
&  CC & HEG & S XVI & Total (XX) & {-2.69$^{+0.24}_{-0.23}$ } &\\
        & &&& 1511 (2000) & -3.17$^{+1.01}_{-0.54}$ &\\
        &  &  &   & 2415 (2001) & -1.00$^{+0.31}_{-0.31}$ &\\
        &&&& 3407  (2001) &-5.34$^{+2.83}_{-2.99}$ &\\
        &&&& 3724 (2002)& -7.19$^{+1.5}_{-0.03}$ &\\
        &&&& 3815 (2003) & -3.54$^{+0.39}_{-0.39}$ &\\
        &&&& 12314 (2010)&  -30 &\\
        &&&& 12472 (2011) & -1.36$^{+0.35}_{-0.97}$ &\\
        &   & MEG & S XVI & Total (XX) & {-2.85$^{+0.34}_{-0.21}$ } &\\
        & &&& 1511 (2000) & -4.66$^{+1.12}_{-1.55}$ &\\
        &  &  &   & 2415 (2001) & -2.29$^{+0.65}_{-0.9}$ &\\
        &&&& 3407  (2001) & -0.42$^{+0.39}_{-0.39}$ &\\
        &&&& 3724 (2002) & -6.98 &\\
        &&&& 3815 (2003)& -3.57$^{+0.44}_{-0.39}$&\\
        &&&& 12314 (2010) & -1302& \\
        &&&& 12472 (2011) & -1.09$^{+0.59}_{-0.59}$ &\\
        
        &TE&HEG&S XVI &Total (XX) & {-3.19$^{+0.34}_{-0.34}$  } &\\ 
        &&&&107 (1999) & -0.94  &\\
        &&&& 2741 (2002)& -10.53$^{+2.53}_{-4.13}$ & \\ 
        &&&& 2742 (2002)& -1.50$^{+1.18}_{1.01}$ &\\ 
        & & & & 2743 (2002)& -1.05$^{+0.92}_{-0.95}$ & \\
        & & & & 3814 (2003)& -4.84$^{+0.31}_{-1.01}$ &\\  
        &&&& 8525 (2008)& -4.11$^{+0.53}_{-0.77}$ &\\
        &&&& 9847 (2008)& -5.42$^{+1.47}_{-1.14}$ &\\
        &&&& 11044 (2010)& -0.76$^{+0.39}_{-0.53}$ &\\
        &&&& 12313 (2010)& -1.54 &\\
        &&&& 13219 (2011)& -0.80$^{+1.58}_{-0.29}$ &\\
        &&&& 16735 (2016)& -2.67$^{+1.24}_{-1.02}$ &\\
        &&MEG&S XVI &Total (XX) & {-2.21$^{+0.25}_{-0.29}$  } & \\ 
        &&&&107 (1999) &  -17.46$^{+9.01}_{-15.64}$ &\\
        &&&& 2741 (2002)& -2.89$^{+1.44}_{-1.11}$ &\\ 
        &&&& 2742 (2002)& -3.06$^{+1.03}_{-2.08}$ &\\ 
        & & & & 2743 (2002)& -1.19$^{+0.44}_{-1.19}$ & \\
        & & & & 3814 (2003)& -4.23$^{+0.47}_{-0.47}$ &\\  
        &&&& 8525 (2008)& -4.20$^{+0.97}_{-0.59}$ &\\
        &&&& 9847 (2008)& -3.18$^{+0.96}_{-0.37}$ &\\
        &&&& 11044 (2010)& -6.63 &\\
        &&&& 12313 (2010)& -8.85 &\\
        &&&& 13219 (2011)& -1.06$^{+1.95}_{-0.12}$ &\\
        &&&& 16735 (2016)& -2.54$^{+0.54}_{-0.49}$ &
\label{t: var}
\end{longtable}
\end{ThreePartTable}




\normalsize

\label{lastpage}
\end{document}